\documentclass[letterpaper, 10 pt, conference]{ieeeconf}  
\usepackage{graphicx}
\usepackage{amsmath}
\usepackage{cite}
\usepackage{amssymb} 
\usepackage{flushend}
\usepackage{xcolor}

\def\MM{\text{MoEMba}}
\def\i{_{i}}

\IEEEoverridecommandlockouts                              

\title{\LARGE \bf
$\MM$: A Mamba-based Mixture of Experts for High-Density EMG-based Hand Gesture Recognition
}

\author{Mehran Shabanpour$^{\dagger}$,  Kasra Rad$^{\dagger}$, Sadaf Khademi$^{\dagger}$, and Arash Mohammadi$^{\dagger}$
\thanks{This work was partially supported by the Natural Sciences and Engineering Research Council (NSERC) of Canada through the NSERC Discovery Grant RGPIN-2023-05654.}
\thanks{$^{\dagger}$All authors are with Concordia Institute for Information Systems Engineering (CIISE), Concordia University, Montreal, Canada.}
}

\begin{document}

\maketitle
\thispagestyle{empty}
\pagestyle{empty}

\begin{abstract}
High-Density surface Electromyography (HD-sEMG) has emerged as a pivotal resource for Human-Computer Interaction (HCI), offering direct insights into muscle activities and motion intentions. However, a significant challenge in practical implementations of HD-sEMG-based models is the low accuracy of inter-session and inter-subject classification. Variability between sessions can reach up to $40\%$ due to the inherent temporal variability of HD-sEMG signals.  Targeting this challenge, the paper introduces the $\MM$ framework, a novel approach leveraging Selective State-Space Models (SSMs) to enhance HD-sEMG-based gesture recognition. The $\MM$ framework captures temporal dependencies and cross-channel interactions through channel attention techniques. Furthermore, wavelet feature modulation is integrated to capture multi-scale temporal and spatial relations, improving signal representation. Experimental results on the CapgMyo HD-sEMG dataset demonstrate that $\MM$ achieves a balanced accuracy of $56.9\%$, outperforming its state-of-the-art counterparts. The proposed framework's robustness to session-to-session variability and its efficient handling of high-dimensional multivariate time series data highlight its potential for advancing HD-sEMG-powered HCI systems.
\newline

\end{abstract}

\section{Introduction}
Surface ElectroMyoGraphy (sEMG) signals~\cite{hodossy2024} are a promising resource for Human-Computer Interaction (HCI),  providing direct insights into muscle activities due to their ability to encapsulate motion intentions~\cite{hu2023}. As such, sEMG-based gesture recognition frameworks emerged as the core technology in developing cutting-edge Muscle-Computer Interfaces (MCIs), facilitating applications ranging from active prostheses~\cite{farina2014, parajuli2019, fleming2021, de2021, lv2023}, wheelchairs~\cite{al2021, iqbal2024} and exoskeletons~\cite{carvalho2023}, as well as in neuromuscular diagnosis~\cite{torres2022}, neurorehabilitation~\cite{eraslan2023}, and video game interactions~\cite{converse2013}. Recent advancements in Artificial Intelligence (AI)-driven solutions have significantly enhanced the development of sEMG-powered MCIs. 
AI-based MCI tasks range from discrete movement classification and joint angle estimation to force/torque estimation offering significant potential to advance EMG pattern recognition. Despite the recent surge of interest, such models fail to address critical challenges faced in non-ideal, practical conditions, including variations in electrode placement and changes in muscle state both across different subjects and within the same person over time~\cite{ni2024, xiong2021}.


\vspace{.025in}
\noindent
\textbf{Literature Review:}
MCI systems commonly rely on a set of standardized functions to represent features from different domains, enhancing the information density embedded in high-density (HD)-sEMG signals, and improving the discrimination of different gestures. The extracted representations can be utilized for sEMG pattern recognition through various classifiers, which are broadly classified into two primary categories: 

\vspace{.025in}
\noindent
\textbf{\textit{Machine Learning (ML)-based Methods~\cite{khushaba2012C1}:}} ML models rely on feature engineering, where relevant features are designed and extracted manually from signals encompassing three primary categories: time-domain (TD), frequency-domain (FD), and time-frequency domain (TFD) features. TD features~\cite{de2021, lv2023} are derived directly from raw sEMG signals, representing their time-dependent variations, and are particularly beneficial due to their low computational complexity. In contrast, FD features~\cite{al2015} are obtained through the Fourier transform of the sEMG signal's autocorrelation function. Building upon these approaches, TFD features~\cite{karheily2022} combine both time and frequency information, providing a comprehensive view of the signal's energy distribution, with wavelet transform being a commonly used technique for this analysis.  Despite the availability of various feature extraction techniques, the multidimensional, non-linear and multichannel nature of sEMG coupled with its inherent non-stationary nature, makes this signal challenging for effective training of ML models~\cite{mian2024}.

\vspace{.025in}
\noindent
\textbf{\textit{Deep Learning (DL)-driven Models~\cite{li2021}:}} DL, on the contrary, is distinguished by its hierarchical architecture. This structure enables the model to progressively extract complex feature representations, automatically capturing underlying patterns and relationships within the data at various levels. The capability of autonomously extracting spatial features made Convolutional Neural Networks (CNNs)~\cite{rahimian2020, tyacke2023} a leading choice for sEMG gesture recognition in DL-driven models. However, limited receptive fields of CNNs restrict their ability to capture long-range dependencies in sEMG signals, which are inherently sequential time series data.
In contrast, Recurrent Neural Networks (RNNs)~\cite{sun2022, jabbari2020}, designed specifically to model temporal relationships and sequential patterns, offer a more suitable approach for analyzing such signals.
Consequently, several approaches utilize hybrid CNN-RNN models~\cite{mian2024, zeng2024}. This combined architecture excels in capturing both spatial and temporal features. On the other hand, transformers offer a compelling alternative with their self-attention mechanism~\cite{zabihi2023, montazerin2023}, granting global receptive fields and enabling superior capture of cross-time dependencies, addressing the limitations of both CNNs and RNNs in this regard. This global perspective, while advantageous, is computationally expensive. The self-attention mechanism necessitates each element attend to all others, leading to computational overhead that scales quadratically with sequence length~\cite{zeng2024, alkilane2024}.

\vspace{.025in}
\noindent
\textbf{Targeted Challenges:} Although DL-driven models have achieved remarkable accuracies in intra-session tasks, their performance diminishes in inter-session and inter-subject scenarios~\cite{pereira2024C1, du2017}.
In the context of sEMG analysis, these terms describe distinct data collection paradigms: intra-session involves data from a single subject during a continuous recording, inter-session refers to recordings from the same subject across multiple sessions, and inter-subject pertains to data from different individuals.

In greater detail, a key challenge in practical implementations is the low accuracy of inter-session and inter-subject classification, with session-to-session variability reaching as high as $10$–$40$\%, which arises from the inherent temporal variability of sEMG signals. This variability is influenced by several factors, including muscle condition (fatigue, atrophy, or hypertrophy), changes in skin impedance (such as sweating), anatomical differences among subjects, non-stationarities due to limb positioning and force variations, and electrode drift, all of which contribute to accuracy degradation. Moreover, these factors can introduce considerable inconsistencies in signal patterns across different sessions and individuals, thus complicating the generalization of gesture recognition algorithms~\cite{li2021, du2017, li2024, islam2024}.

The challenge of between-session distribution shifts has been conventionally handled through experimental design by developing multi-user and multi-session training protocols, as well as data processing techniques such as data augmentation~\cite{du2017, li2024}, unsupervised learning, and transfer learning~\cite{xia2024, emimal2024} with advanced DL architectures. Despite offering improvements in accuracy, these methods often require large amounts of adaptation data~\cite{alkilane2024, li2024, pereira2024C2}. Moreover, sEMG data collection is a time-consuming and resource-intensive process~\cite{du2017, pereira2024C2}, requiring specialized equipment and trained personnel, which further limits the availability of data. Moreover, the significant computational demands of these methods restrict their feasibility for real-time applications, particularly in resource-constrained embedded systems~\cite{alkilane2024, li2024}.

A key consideration, often overlooked in most above-mentioned models, is the inter-channel dependencies inherent in sEMG data. Gesture movements involve complex inter-muscular coordination, which is reflected in the inter-channel relationships captured by the placement of sEMG electrodes. Many models incorrectly apply a channel-independent approach, treating the data as uncorrelated univariate time series and thus neglecting these essential inter-channel relationships~\cite{zeng2024, alkilane2024, xia2024}. Although some approaches~\cite{liu2023, nie2022} attempt to address this limitation by combining channels using mechanisms such as self-attention, linear combinations, or convolutions, these methods are computationally expensive and generally miss proportional relationships by modeling inter-channel relationships as weighted sums.

\vspace{.025in}
\noindent
\textbf{Contributions:}
Recognizing the crucial role of cross-channel dependencies in multivariate time series, we aim to capture temporal dependencies in HD-sEMG signals using Selective State Spaces (SSS), and address cross-channel interactions through techniques of channel attention. Recent advancements in State-Space Models (SSMs)~\cite{alkilane2024}, notably Mamba, offer efficient modeling of sequential data dynamics with linear computational complexity without losing the global receptive
field, even for long sequences, positioning them as a potential alternative to transformers. In other words, SSMs adopt an RNN-like approach that facilitate dynamic adaptation to data distribution shifts, enabling the capture of diverse temporal features such as long-term trends and short-term fluctuations.

The paper introduces the $\MM$ framework built upon the Mamba architecture—a novel lightweight advanced SSM-based approach—specifically designed to address the critical challenge of session-to-session variability while prioritizing simplicity and efficiency. In summary, the paper makes the following contributions:
\begin{itemize}
\item As the first application of Mamba for HD-sEMG hand gesture recognition, $\MM$ model uses an adaptive combination of multiple Mamba experts within a Mixture of Experts (MoE) configuration to capture both short-term and long-term gesture dynamics. 
\item Wavelet Transform Feature Modulation (WTFM) is incorporated to capture multi-scale temporal and between-channel spatial relations, integrating both time-domain and frequency-domain information to enhance signal representation.
\item The $\MM$'s design, including its architecture and feature extraction, is  computationally efficient requiring fewer Floating Point Operations per Second (FLOPS) than State-Of-The-Art (SOTA) models. This efficiency, combined with robustness to session-to-session variability in HD-sEMG recordings makes $\MM$ ideal for real-world applications like prosthetic control and human-computer interaction. 
\end{itemize}
The remainder of the paper is organized as follows: Section~\ref{sec:met} begins by describing the dataset used in this research and provides an overview of the relevant background. Section~\ref{sec:Xmamba} introduces our proposed framework, $\MM$, detailing its architecture and key components. Section~\ref{sec:res} presents a comprehensive analysis of the experimental results. Finally, Section~\ref{sec:con} concludes the paper.

\section{Materials and Methods}\label{sec:met}
In this section, first, we briefly present the HD-sEMG dataset used for development and testing of the $\MM$ framework. Then, required background on State-Space Models (SSMs) is introduced.

\vspace{.3in}
\noindent
\textit{A. Dataset}

In this study, we used the CapgMyo HD-sEMG dataset~\cite{du2017}, recorded from $128$ channels sampled at $1000$Hz. We utilized sub-databases DB-a ($23$ subjects performing $8$ gestures held for $3-10$s) and DB-b (a subset of DB-a comprising $10$ subjects with recordings from two sessions separated by a minimum of $7$ days, performing the same $8$ gestures with hold durations of $\sim$$3$s).  The CapgMyo HD-sEMG dataset configuration introduces inter-session and inter-subject variability, which challenge model robustness—an issue that this study aims to address. To capture richer semantic information and local features, the data was pre-processed using $45-55$Hz Butterworth filter, followed by segmentation into overlapping $64$ms windows with $8$ms steps, resulting in $64\times128$ matrices, and finally normalized to $[-1, 1]$. 

\vspace{.025in}
\noindent
\textit{B. Selective State-Space Models}

The SSMs are mathematical models used to represent dynamic systems through state variables~\cite{gu2021}. Fundamentally, SSMs characterize the evolution of a system through two primary equations, i.e., the state model, and the observation model. The state model defines the evolution of the hidden state $h(t)$ over time as influenced by the input $x(t)$
\begin{eqnarray}
h'(t) = Ah(t) + Bx(t), 
\end{eqnarray}
where \( A \in \mathbb{R}^{N \times N} \) is the state transition matrix, \( B \in \mathbb{R}^{N \times 1} \) is the input matrix, \( h(t) \in \mathbb{R}^{N} \) is the hidden state at time $t$, and $h'(t)$ is the derivative of the hidden state. This evolution of the hidden state is then used by the observation equation to determine the output $y(t)$
\begin{eqnarray}
y(t) = Ch(t) + Dx(t), 
\end{eqnarray}
where \( C \in \mathbb{R}^{1 \times N} \) is the output matrix, and \( D \in \mathbb{R} \) is the direct feedthrough term (often set to zero).
The application of SSMs within ML context requires the discretization of the continuous-time formulations. A common discretization method is the Zero-Order Hold (ZOH), which assumes a constant function value over an interval.
After ZOH discretization, the SSM equations can be rewritten as follows
\begin{eqnarray}
h_k &=& \overline{\mathbf{A}}h_{k-1} + \overline{\mathbf{B}}x_k, \\
y_k &=& \mathbf{C}h_k, 
\end{eqnarray}
where $\overline{\mathbf{A}} = \exp(\Delta \mathbf{A})$, and $\overline{\mathbf{B}} = (\Delta \mathbf{A})^{-1}(\exp(\Delta \mathbf{A}) - \mathbf{I})$. Furthermore,   $h_k$ is the hidden state at discrete time step $k$, $x_k$ and $y_k$ represent the input and output at discrete time step $k$, and $\Delta = [t_{k-1}, t_k]$.
However, a key limitation of traditional SSMs lies in their Linear Time-Invariant (LTI) nature, where fixed parameters such as $A, B, C$, and \( \Delta \) restrict their ability to adapt to diverse sequences. Mamba~\cite{gu2023}, a recent model developed based on SSMs, addresses this issue by introducing a selection mechanism that parameterizes these matrices as functions of the input \( x \), enabling input-dependent dynamics. The parameterization is defined as 
\begin{eqnarray}
B \rightarrow S^B &=& W^B x, \label{(8)} \\
C \rightarrow S^C &=& W^C x, \label{(9)}\\
\Delta \rightarrow S^\Delta &=& \tau_\Delta \cdot \text{BroadCast}_D(W^\Delta x), \label{(10)}
\end{eqnarray}
where \( W^B \), \( W^C \), and \( W^\Delta \) are learnable projection matrices, \( \tau_\Delta \) is the softplus activation function, and ${BroadCast}_D$ is a function that replicates the result of $W^\Delta x$ across all feature dimensions. This transition to a time-variant model allows Mamba to adaptively filter irrelevant information while retaining critical context, making it highly effective for tasks requiring long-context modeling ~\cite{qu2024}.

\section{The $\MM$ Framework}\label{sec:Xmamba}
\setlength{\textfloatsep}{0pt}
\begin{figure*}[t]
      \centering
      \includegraphics[scale=0.40]{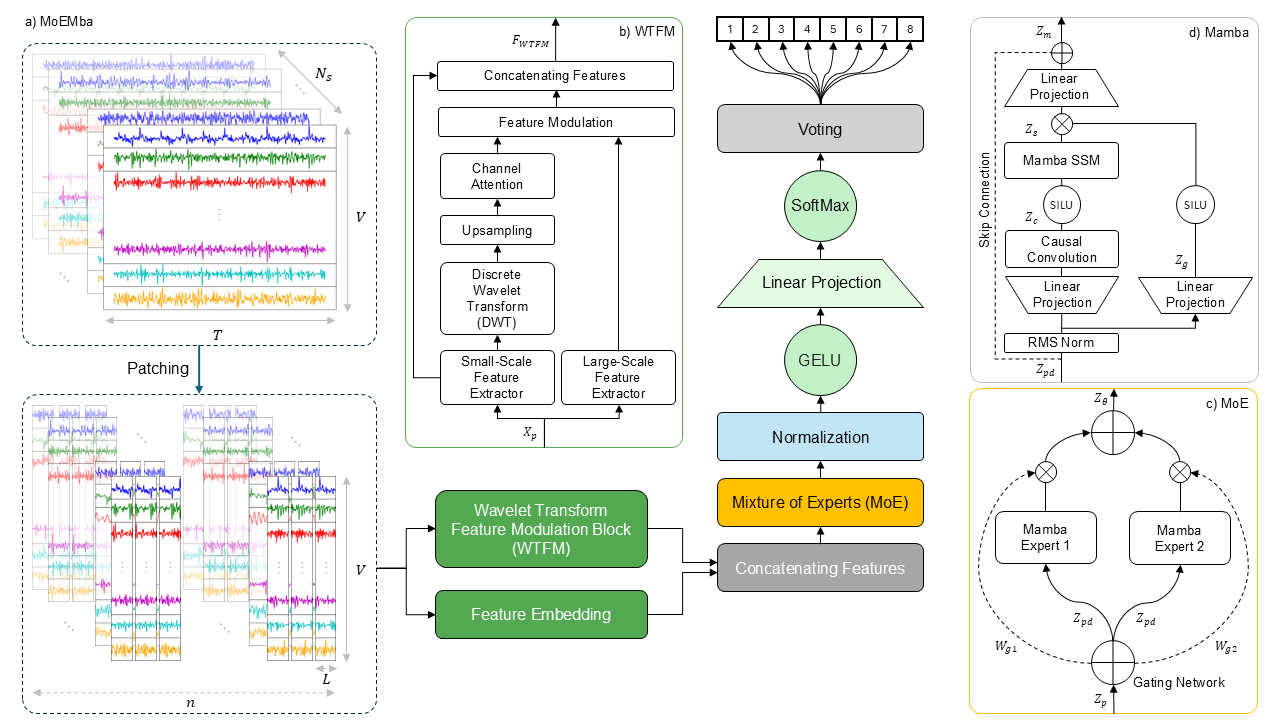}
         \vspace{-.1in}
      \caption{ $\MM$ Framework. (a) Raw time series data is pre-processed via patching. (b) Shallow features are extracted using small $(3\times3)$ and large $(7\times7)$ receptive field convolutions, and wavelet transform-modulated features, with channel attention applied to EMG signal patches. (c) A Mixture of Experts (MoE) block routes patches to two Mamba experts based on a gating network. (d) The Mamba block incorporates projections, $1$D convolution, selective State-Space Modeling (SSM), and skip connections for temporal dependency learning.}
      \label{fig1}
      \vspace{-.25in}
\end{figure*}
The proposed $\MM$ framework is built based on the Mamba architecture, chosen for its efficient inference and ability to handle long sequences in light of its linear scaling with sequence length. 
$\MM$ decodes muscle activity and recognizes gestures by establishing contextual dependencies across EMG signal channels. As depicted in Fig.~\ref{fig1}, the architecture comprises of three key modules: \textit{(i) WTFM Block:} This block is employed as a shallow feature extraction module to enhance multi-scale temporal representations; \textit{(ii) MoE Block}: This block combines Mamba-based experts via sparse gating for dynamic pattern modelling; and \textit{(iii) Classification Block}: A voting mechanism is used in the classification block to ensure robust signal-level predictions. Detailed description of each block is presented in the subsequent sub-sections.

\vspace{.025in}
\noindent
\textit{A. The WTFM Block}

In multivariate time series classification, particularly when dealing with high-dimensional data such as HD-sEMG signals, capturing both local temporal dynamics and global contextual information is crucial. In this regard and inspired by a recent work~\cite{huang2024}, we adopt a multi-scale representation learning through wavelet transform feature modulation. Such an approach enhances feature representation learning from HD-sEMG signals.

The $\MM$ processes each signal patch, $P_i \in \mathbb{R}^{L \times V}$, by the WTFM block, where $L$ defines the temporal length and $V$ denotes the number of channels. This $2$D representation facilitates convolutional operations, allowing a more effective capture of spatial relationships across channels. The WTFM block initially extracts local and global temporal features using small (e.g., $3\times3$) and large (e.g., $7\times7$) temporal convolutional kernels, respectively. A Discrete Wavelet Transform (DWT) is then applied to the local temporal features (extracted by the small temporal convolutional kernel), decomposing them into four directional components: approximation $(c_A)$, horizontal detail $(c_H)$, vertical detail $(c_V)$, and diagonal detail $(c_D)$. These components are calculated based on different frequency bands of the wavelet decomposition. 
These components capture detailed temporal variations at different scales. DWT components are then upsampled back to the original size of $P_i$ using bicubic interpolation. These upsampled wavelet features are then refined using a channel attention mechanism, which learns to focus on the most important channels. The attention weights $\alpha\i$ are given by
\begin{eqnarray}
\alpha\i = \sigma(W_2 \cdot \text{ReLU}(W_1 \cdot \text{MaxPool}(c^i_{*}))), 
\end{eqnarray}
where $W_1$ and $W_2$ are learnable parameters, $c^i_{*}$ refers to wavelet components ($c^i_A, c^i_H, c^i_V$, and $c^i_D$) associated with $i^{\text{th}}$ patch. MaxPool($\cdot$) reduces the spatial dimensions, and $\sigma(\cdot)$ is the sigmoid function. The enhanced wavelet features are then combined with the original large-scale features (extracted by the large temporal convolutional kernel) using a Hadamard product to capture both global and local information. This modulation process enhances the model's ability to discriminate subtle temporal patterns within the EMG signals while retaining crucial global information. After modulation, the refined large-scale features are concatenated with the small-scale features to integrate both high-level temporal representations and fine-grained local details. This fusion enables the model to utilize complementary information from different temporal scales, further improving its ability to capture intricate sEMG signal patterns.
Finally, the combined feature set ($F_{\text{WTFM}}$) is fused with raw patch embeddings ($F_{\text{chan}}$)—obtained via 1D convolutional value embedding function encapsulating the original signal's channel characteristics—to integrate both FD and TD information, improving encoder performance. The model backbone network then processes the fused features $Z_{\text{p}}$ for deeper feature extraction.
\begin{eqnarray}
Z_{\text{p}} = \text{Concat}(F_{\text{WTFM}}, F_{\text{chan}}), \text{ where } Z_{\text{p}} \in \mathbb{R}^{N_s\times n \times E \times V} 
\end{eqnarray}
This completes description of the WTFM block of the proposed $\MM$ framework. Next, we present the MoE module of the proposed $\MM$ framework.

\vspace{.025in}
\noindent
\textit{B. The MoE Block}

The MoE block is utilized with each expert being implemented as a Mamba module.
The MoE framework consists of a set of $\eta$ experts ($E_1, \dots, E_\eta$), each specializing in learning distinct patterns within the time series data, and a gating network ($G$) that dynamically computes a sparse $\eta$-dimensional vector to weight the contributions of each expert. Given a sequence of embedded patches $Z_p$, the final output $Z_\theta$ is computed from the expert outputs as follows
\begin{eqnarray}
Z_\theta = \sum_{i=1}^{\eta} G_i(Z_p) \cdot E_i(Z_p).  
\end{eqnarray}
This formulation allows the MoE framework to adaptively combine the strengths of multiple Mamba experts, facilitating robust modeling of both short-term and long-term trends in hand gesture data. 

We incorporate an adapted gating mechanism inspired by~\cite{alkilane2024} that introduces optimizing enhanced sparsity to optimize computational efficiency of the underlying model. More specifically, a tunable sparsity mechanism is employed where Gaussian noise, controlled by a trainable weight matrix $W_{\text{noise}}$, is added to the network output before applying the Softmax function. To further promote sparsity, only the top $k$ values from the network output are retained, while the remaining values are suppressed to $-\infty$, effectively setting their corresponding gate values to zero. This selective approach is formalized as follows
\begin{eqnarray}
\!\!\!\!\!\!\!\!\mathcal{G}(Z_p) &\!\!\!\!=\!\!\!\!& \text{Softmax}(\text{TopKGating}(H(Z_p), k)), \\
\!\!\!\!\!\!\!\!\text{where } H_i(Z_p) &\!\!\!\!=\!\!\!\!& (Z_p W_{\text{gate}})_i + \text{Softplus}((Z_p W_{\text{noise}})_i),
\end{eqnarray}
and $\text{TopKGating}(v, k)_i$ retains only the top $k$ elements of $v$, setting the rest to $-\infty$. Here, $W_{\text{gate}}$ and $W_{\text{noise}}$ are trainable weight matrices. 
To prevent the potential bias in the gating network towards a single expert, a balanced gate mechanism is implemented. This is achieved by minimizing a balance loss $\mathcal{L}_B$, which encourages a more uniform distribution of samples across experts. The balance loss is defined as
\begin{eqnarray}
\mathcal{L}_B(Z_p) = \lambda_B \cdot \text{CV}(\text{Load}(Z_p))^2, 
\end{eqnarray}
where $\text{CV}(\cdot)$ is the coefficient of variation of the load vector, the Load function ($\text{Load}(\cdot)$) is a smooth estimator used to quantify the distribution of data samples across the different experts, and \(\lambda_B\) is a scaling factor. Additionally, a regularization term \(\mathcal{L}_Z\) has been used to penalize large logits within the network, further stabilizing the gating mechanism
\begin{eqnarray}
\mathcal{L}_Z(Z_p) = \frac{1}{n} \sum_{i=1}^n \left( \log \sum_{j=1}^\eta e^{(Z_p)_j} \right)^2,
\end{eqnarray}
where \(n\) is the number of patches and \(\eta\) is the number of experts. The total auxiliary loss \(\mathcal{L}_{\text{aux}}\) is a weighted combination of these losses, integrated into the overall model loss function to ensure balanced and efficient expert utilization. Such a gating network allows  to capture complex temporal dependencies in HD-sEMG signals.

\begin{figure*}[t]
\centering
\includegraphics[scale=0.35]{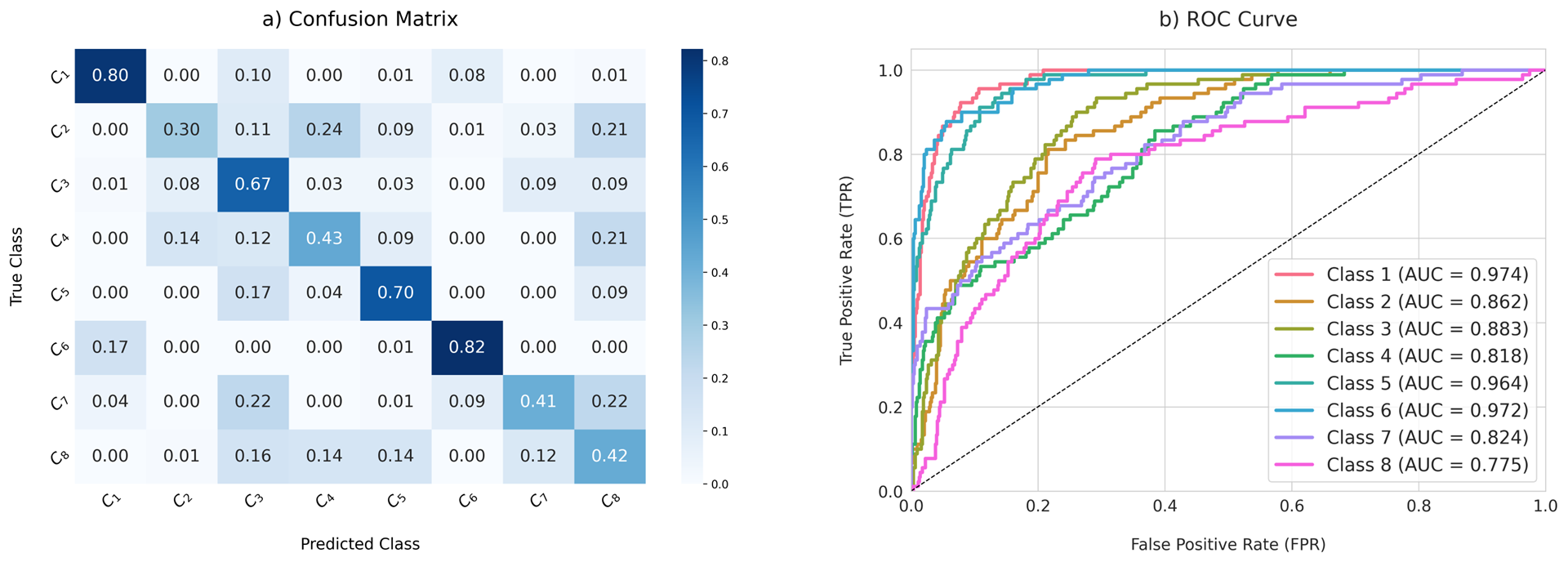}
\vspace{-.3in}
\caption{Confusion Matrix and Receiver Operating Characteristic (ROC) Curves. (a) The heatmap visually represents the class-wise prediction probabilities within the primary dataset. The horizontal axis corresponds to predicted labels, while the vertical axis indicates true labels. Diagonal values reflect individual class accuracies, whereas off-diagonal values represent misclassification probabilities. Darker colors signify higher probabilities, while lighter shades indicate lower ones. The chance level is $0.125$. (b) The ROC curves illustrate the classification performance of $\MM$ across different discrimination thresholds. Each class is color-coded, with corresponding labels and the Area Under the ROC Curves (AUC) provided in the legend at the bottom right. The dashed black diagonal line represents random chance classification.}
      \label{fig2}
      \vspace{-.25in}
\end{figure*}

\vspace{.025in}
\noindent
\textit{C. Classification Block}

The final stage of the $\MM$  uses a classification head to process the MoE output, normalizing and activating it prior to projection into the class space. The resulting logits are then transformed into probabilities via softmax, and the predicted class is determined by selecting the highest probability. Afterwards, predictions from patches belonging to the same signal are aggregated, and majority voting is used to determine the final classification of the entire signal. This approach reduces the impact of noisy/inconsistent predictions at the patch level.

\section{Experimental Results}\label{sec:res}
In this section, we evaluate performance of the proposed $\MM$ framework on the CapgMyo DB-b HD-sEMG datasets~\cite{du2017}, and provide comparisons with baseline SOTA models. The primary objective is addressing the challenge of inter-session variability to have  more consistent and generalizable performance, both within and between subjects. Single-trial EMG signals inherently suffer from a low Signal-to-Noise Ratio (SNR) and variability due to factors such as electrode placement, skin impedance, time of day, and participant conditions. Consequently, we evaluated the inter-session gesture recognition performance on the DB-b dataset, utilizing a train-on-one-session, test-on-the-other approach, while the DB-a dataset was incorporated for data augmentation during the training phase. However, due to data corruption in the final subject’s recordings, the experiment proceeded with the first nine subjects in DB-b~\cite{pereira2024C1}. 
The model's gesture classification performance was then evaluated across eight distinct classes: ``Thumb up," ``Extension of index and middle with flexion of the others," ``Flexion of ring and little finger with extension of the others," ``Thumb opposing the base of the little finger," ``Abduction of all fingers," ``Fingers flexed together in a fist," ``Pointing index," and ``Adduction of extended fingers".

All experiments were conducted on a computing system equipped with an NVIDIA GeForce RTX $2080$ Ti GPU, an Intel i$9$-$9820$X CPU, and $128$GB of RAM. The $\MM$ framework was configured with several key hyperparameters. The SSM employed a state expansion factor of $16$ and a local convolution width of $4$, with a model dimension of $128$. The MoE architecture consisted of two experts and two gating networks, with a hidden dimension factor of $1$. Each Mamba expert within the MoE framework used a block expansion factor of $4$.  A small number of experts was chosen to reduce the risk of overfitting, particularly given the dataset size, while ensuring that each expert receives sufficient data for effective training. The model was trained over $50$ epochs, using a batch size of $32$ and an initial learning rate of $0.0001$, which was adjusted via a cosine annealing schedule.

Performance evaluation was conducted using three well-established classification metrics: total accuracy, confusion matrix, and Receiver Operating Characteristic (ROC) curves. Total accuracy was computed using a ``winner-take-all" approach, measuring the percentage of correctly assigned labels. The confusion matrix provided a class-wise breakdown of correct classifications and false positive/negative rates. Given the eight-class setup, the resulting confusion matrix (Fig.~\ref{fig2} (a)) was $8\times8$, where diagonal elements indicated correct classifications, while off-diagonal elements represented misclassifications. Additionally, the ROC curve (Fig.~\ref{fig2} (b)) is used to assess classification performance across varying thresholds, with the Area Under the Curve (AUC) serving as a summary measure of classifier effectiveness.

\renewcommand{\arraystretch}{1.5}
\begin{table}[t!]
\caption{Comparison results using the DB-b dataset.}
\vspace{-.2in}
\label{table_acc}
\begin{center}
\begin{tabular}{|c|c|c|}
\hline
\textbf{Model} & \textbf{Author} & \textbf{Accuracy}\\
\hline
TD & K. Englehart et al.~\cite{englehart2003} & $0.416$±$0.198$\\
\hline
ETD & R. N. Khushaba et al.~\cite{khushaba2012C2} & $0.437$±$0.209$\\
\hline
NinaPro & M. Atzori et al.~\cite{atzori2014} & $0.425$±$0.170$\\
\hline
SampEn & A. Phinyomark et al.~\cite{phinyomark2013} & $0.448$±$0.217$\\
\hline
TVGGNet & J. Pereira et al.~\cite{pereira2024C1} & $0.420$±$0.183$\\
\hline
The Proposed $\MM$ & This Work & $0.569$±$0.201$\\
\hline
\end{tabular}
\end{center}
\end{table}

Table~\ref{table_acc} summarizes the performance metrics of $\MM$ and its counterpart models. As the results highlight, $\MM$ achieved a balanced accuracy of $56.9\%$, which is significantly above the random chance level of $12.5\%$ for an eight-class problem. Notably, model comparisons revealed a trade-off between accuracy and network complexity, with the $\MM$ framework achieving performance comparable to SOTA transformer-based architectures while maintaining substantially lower computational complexity. Unlike transformer models, which typically have quadratic complexity $O(n^2)$, $\MM$ operates with linear complexity $O(n)$, making it well-suited for long time-series processing, such as sEMG signal classification. 

The ROC analysis further illustrated class-specific performance, with AUC values of $0.974$, $0.862$, $0.883$, $0.818$, $0.964$, $0.972$, $0.824$, and $0.775$ across the eight gesture classes. The variation in classification accuracy can be attributed to differences in muscle activation patterns. Gestures such as ``Thumb up" and ``Abduction of all fingers" exhibited higher accuracy due to their distinct and easily recognizable muscle activation, whereas gestures involving partial flexion and extension, such as ``Extension of index and middle with flexion of the others" and ``Adduction of extended fingers," were more prone to misclassification due to overlapping signal features. Moreover, inter-session variability, particularly changes in electrode placement and skin conditions, posed additional challenges in maintaining classification consistency.

Beyond accuracy, the computational efficiency of $\MM$ underscores its potential for real-time applications. FLOPS is a key metric that quantifies the number of floating-point operations a system can perform per second, providing insight into the processing power required. The model’s lightweight architecture and compact design, with a total of $455,003$ parameters and $27,312,144$ FLOPS, enable efficient processing of long sequences while maintaining implementation feasibility. Furthermore, its streamlined architecture ensures scalability, allowing deployment across various platforms without significant computational overhead. However, certain limitations remain, such as the need for larger training datasets to enhance generalization and the risk of overfitting when applied to smaller datasets.

\section{Conclusion}\label{sec:con}
In this study, we introduced the $\MM$ framework, a novel approach for HD-sEMG-based hand gesture recognition, leveraging the Mamba architecture to address the critical challenge of session-to-session variability. Our experimental results demonstrated that the $\MM$ framework achieves superior performance compared to its SOTA counterparts, with a balanced accuracy of $56.9\%$ on the CapgMyo DB-b dataset. This performance is achieved while maintaining computational efficiency. The integration of wavelet feature modulation and channel attention mechanisms within the $\MM$ framework enhances the model's ability to capture both local and global temporal dependencies, leading to improved gesture recognition accuracy. Additionally, the MoE configuration allows the model to dynamically adapt to varying patterns in the HD-sEMG signals.
A fruitful direction for future research is optimizing the model's robustness to inter-session variability by exploring synthetic data to augment real-world datasets. Another direction for future research is to investigate the potential of incorporating additional modalities, to further improve the accuracy and reliability of gesture recognition systems.

\bibliographystyle{IEEEtran}
\footnotesize
\bibliography{refs}

\end{document}